\documentclass[aps,twocolumn,showpacs,lengthcheck,preprintnumbers]{revtex4-1} 
\usepackage{amsfonts}
\usepackage{amsmath}
\usepackage{amssymb}
\usepackage{graphicx}
\usepackage{fontenc}
\usepackage{color}

\usepackage[unicode]{hyperref}
\usepackage{microtype}
\hypersetup{
	pdftitle={ultracoldgas pendulum},
	pdfauthor={},
	pdfsubject={},
	colorlinks=true,
	linkcolor=blue,
	citecolor=blue,
	filecolor=black,
	urlcolor=blue
}

\begin{document}
	
	\title{Superfluid rings as quantum pendulums}

\author{Antonio Mu\~noz Mateo}
\affiliation{Departamento de F\'isica, Universidad de La Laguna, 38200 Tenerife, Spain}
\author{Grigory E. Astrakharchik}
\affiliation{Departament de F\'isica, Universitat Polit\`ecnica de Catalunya, 08034 Barcelona, Spain}
\affiliation{Departament de F\'isica Qu\`antica i Astrof\'isica, Universitat de Barcelona, Mart\'i Franqu\`es 1, 08028 Barcelona, Spain.}
\affiliation{Institut de Ci\`encies del Cosmos (ICCUB), Universitat de Barcelona, Mart\'i Franqu\`es, 1, 08028 Barcelona, Spain.}
\author{Bruno Juli\'{a}-D\'{i}az}
\affiliation{Departament de F\'isica Qu\`antica i Astrof\'isica, Universitat de Barcelona, Mart\'i Franqu\`es 1, 08028 Barcelona, Spain.}
\affiliation{Institut de Ci\`encies del Cosmos (ICCUB), Universitat de Barcelona, Mart\'i Franqu\`es, 1, 08028 Barcelona, Spain.}

\begin{abstract}
A feasible experimental proposal to realize a non-dispersive quantum pendulum is presented. 
The proposed setup consists of an ultracold atomic cloud, featuring attractive interatomic interactions, loaded into a tilted ring potential. 
The classical and quantum domains are switched on by tuned interactions, and the classical dynamical stabilization of unstable states, i.e. {\it a la} Kapitza, is shown to be driven by quantum phase imprinting. The potential use of this system as a gravimeter is discussed.
\end{abstract}

\maketitle

\begin{figure*}[tb]
\includegraphics[width=0.8\linewidth]{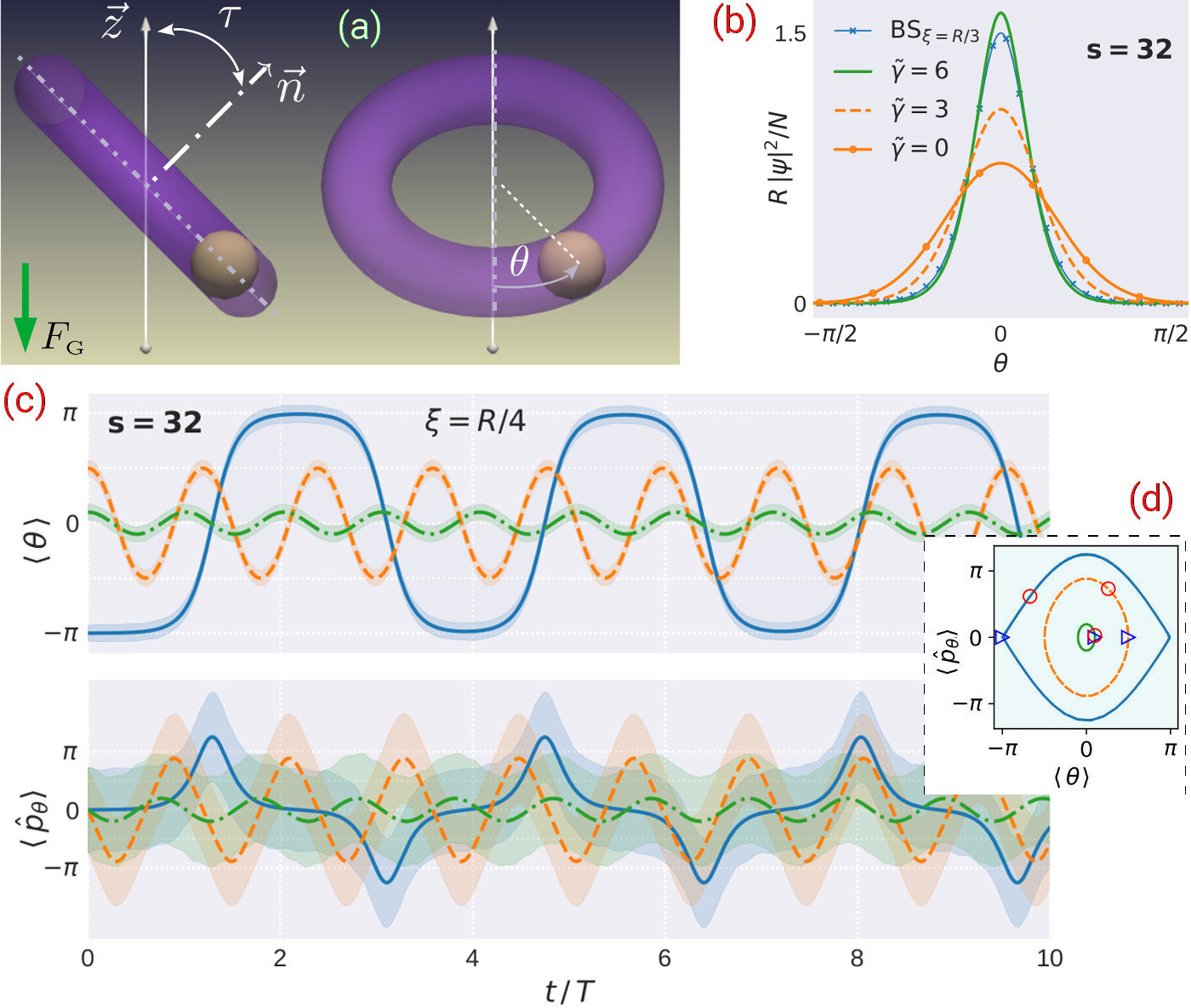}
\caption{(a) Schematic picture, side (left) and front (right) views, of a tilted ring potential (purple isocontour) inside which a BEC (cream blob) moves in the presence of the constant gravitational force $F_{\mbox{\tiny G}}$.
(b) Ground-state density profile in a ring trap, characterized by the characteristic frequency $\omega=2\hbar/mR^2$ and potential depth $s=32$ (see text), for various strengths of the interparticle attraction, parameterized by $\tilde \gamma=|\gamma| m N R/\hbar^2$. (c) Oscillations of the center of mass (top panel) and the average momentum (bottom panel) of a bright soliton of width $\xi=R/4$ (or, equivalently, $\tilde\gamma=8$) for different initial positions. The shaded areas around the solid lines represent the mean width in position $\sigma$ and momentum $\sigma_k$ spaces. (d) Trajectories in the phase space. States at $t=0$ (triangles)
and $t=T$ (circles) are indicated by open symbols.
}
\label{fig:oscillations}
\end{figure*}

Ring geometries are omnipresent in physics. Mathematically, they endow systems with periodic boundary conditions; physically, they realize the minimal block of cyclic transport, which would become perpetual if there were no dissipation. Approaching the dissipationless limit, superconductors and superfluids are capable of making the cyclic transport of charge or particles, if not perpetual at least persistent, a particularly striking demonstration of which is the persistent flow of superconducting gravimeters~\cite{Geophysics}. In this regard, the first realizations of ring geometries in ultracold gases opened new avenues for experiments with persistent currents of highly controlled Bose-Einstein condensates~\cite{Ryu2007,Heathcote2008,Ramanathan2011,Moulder2012}. However, in this case, the gravitational pull is an apparent hindrance to stationary flows, since a small tilting of the ring axis with respect to the vertical direction gives rise to an unwanted azimuthal potential for the trapped atoms~\cite{Heathcote2008,Pandey2019}. Interestingly, although suppression of the tilting is necessary for the study of unobstructed currents, by letting the tilting occur the ring is transformed into a pendulum (see Fig.~\ref{fig:oscillations}); for instance, horizontally setting the axis of a typical ring of radius $R=20\,\mu$m converts it into a pendulum of angular frequency $\omega=\sqrt{g/R}=700$~rad/s, where $g$ is the gravity acceleration. While for a repulsively interacting, extended Bose-Einstein Condensate (BEC) the resemblance to the classical pendulum would only apply to the motion of the center of mass, one expects to find its quantum analogue in attractively interacting, localized BECs. The present work is devoted to exploring the validity of this analogy.

The pendulum dynamics has been addressed under multiple perspectives in the 
context of ultracold atomic systems. Its presence is implicitly evident in the 
coherent tunneling of particles, as observed in the Josephson effect~\cite{Smerzi1997,Fialko2015,Pigneur2018}. Additionally, the proposal for the dynamic 
generation of nonlinear excitations has shed further light on the intricacies of 
pendulum dynamics in this context~\cite{Fialko2012}. Most of the studies have focused 
on the dynamical stabilization of pendulum-like equilibrium states in optical 
potentials that are unstable~\cite{Gilary2003,Jiang2021,He2023}.  Closer to our discussion on dynamical stabilization, Ref. \cite{Abdullaev2003} addressed the general dynamics of bright solitons in periodically, rapidly-varying time traps. In a classical 
context, the pendulum dynamics of a microscopic colloidal particle in a ring built 
with optical tweezers has been recently reported~\cite{Richards2018}.
		
{\bf The tilted ring}. We assume a quasi-one-dimensional character of the system, 
which can be ensured by imposing tight transverse confinement to the atoms. 
While the role of quantum fluctuations is enhanced in one dimension, still, 
in the mean-field regime, the gas is sufficiently coherent to be correctly 
described by the Gross-Pitaevskii equation. Although strictly speaking 
Bose-Einstein condensation is absent in one dimension, properties of a 
quasicondensate are correctly captured by equations describing the evolution of 
the condensate wave function. 
A tilting angle $\tau\in[0,\,\pi/2]$ produces on the particles of mass $m$ the 
gravitational potential $U(\tau,\theta)=-m\,g\,R\sin\tau\,\cos\theta$ along 
the azimuthal coordinate $\theta\in[-\pi,\,\pi]$ of the ring, or alternatively 
$U(\tau,\theta)=-m\omega^2 R^2\cos\theta$, where $\omega=\sqrt{g\sin\tau/R}$ is 
the pendulum angular frequency, so $T=2\pi/\omega$ is the corresponding time 
period, see the depiction in Fig.~\ref{fig:oscillations}. 
For non-interacting particles, 
the time evolution is described by the Schr\"odinger equation, i.e. Eq.~\ref{eq:GP1D} with $\gamma=0$ (see the early discussion of pendulums in quantum mechanics by Condon~\cite{Condon1928}, or for a more recent account, for instance, Ref. \cite{Doncheski2003}).
It admits general solutions that can be written as superpositions 
of Mathieu functions~\cite{NIST:DLMF} ${\rm ce}_{2n}(\theta,q)$ and 
${\rm se}_{2n+1}(\theta,q)$, with $n=0,1,2,\dots$, and 
$2q=-(2R/a_{\rm ho})^4$. Here we have introduced the harmonic 
oscillator length, $a_{\rm ho}=\sqrt{\hbar/m\omega}$, associated with 
small-amplitude oscillations around the equilibrium point $\theta=0$. 
For $q\rightarrow 0$, that is for a small tilting or 
$\omega\rightarrow 0$, these functions tend to the trigonometric 
functions $\cos(2n\theta)$ and $\sin[(2n+1)\theta]$.

\begin{figure}[tb]
\flushleft ({\bf a})\\
\centering
\includegraphics[width=\linewidth]{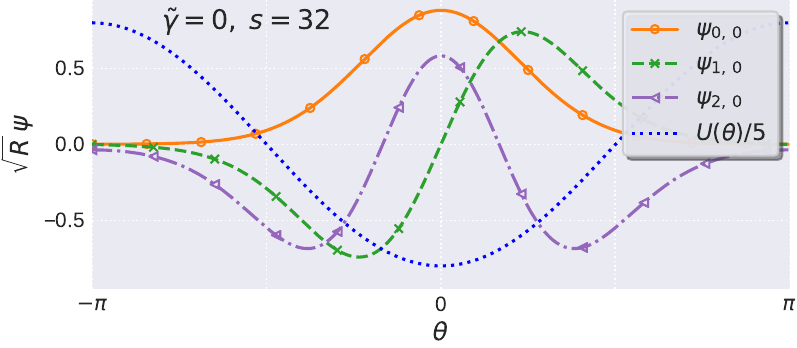}\\
\vspace{-0.6cm}
\includegraphics[width=\linewidth]{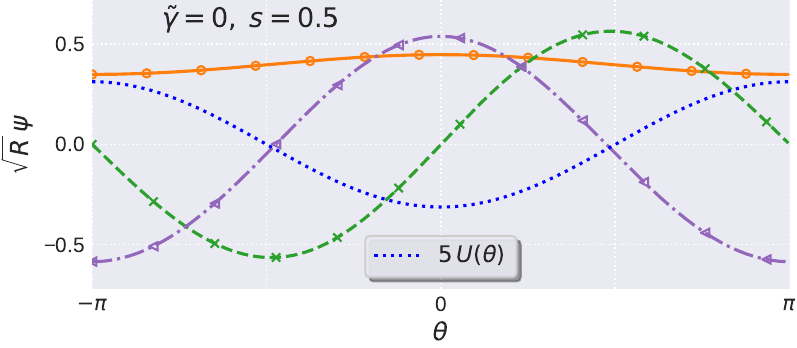}\\ \vspace{-0.4cm}
\flushleft ({\bf b})
\vspace{-0.2cm}
\includegraphics[width=\linewidth]{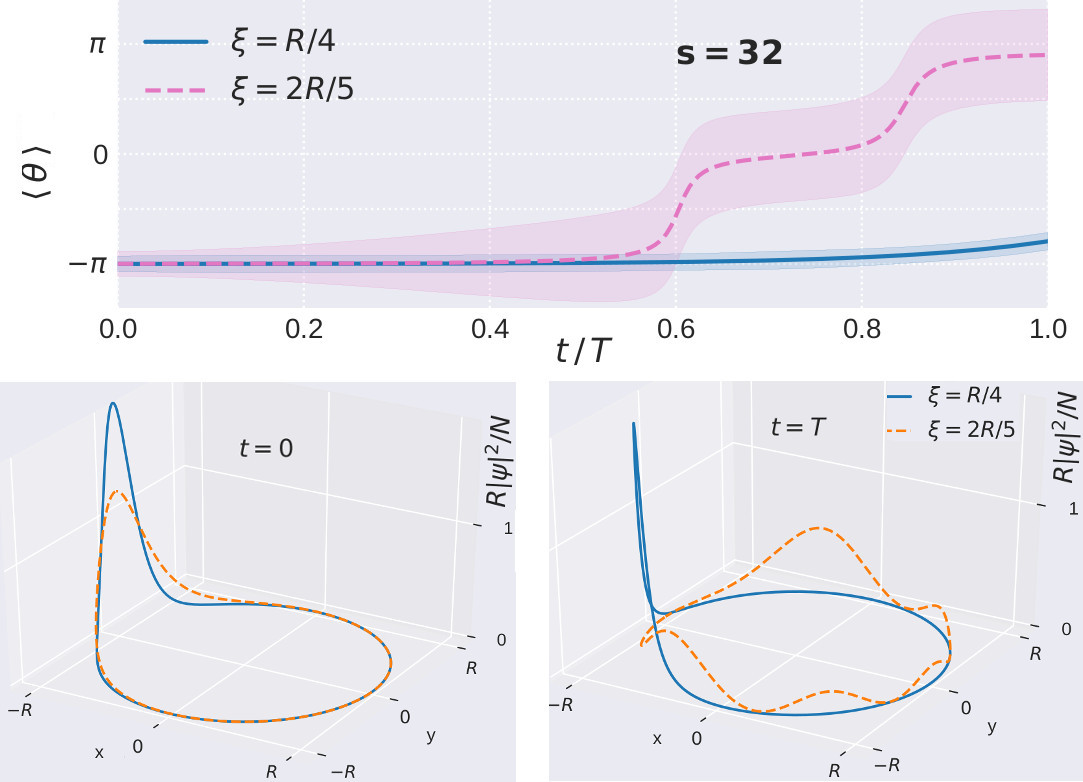}
\caption{ (a) Lowest energy eigenstates (Mathieu functions) of noninteracting systems 
 with potential depths $s=32$ (top panel) and $s=0.5$ 
(bottom panel). (b) Dynamics dependence on the width $\xi$ of an initial 
wave packet (a free soliton) centered close to the classical unstable 
equilibrium point $\theta=\pi$ [in cartesian coordinates $(x,y)=(-R,0)$]. The top panel shows the 
time evolution of the center of mass for $\xi=R/4$ (or interaction 
parameter $\tilde\gamma=8$) and $\xi=2R/5$ ($\tilde\gamma=5$). While the former 
case, well beyond the threshold width for the existence of equilibrium 
states at $\theta=\pi$, follows the classical path  (the same shown 
in Fig.~\ref{fig:oscillations} for much longer evolution), the wider 
soliton spreads over the ring (bottom panels).}
		\label{fig:linear}
	\end{figure}

Deeper insights can be drawn by considering the particle positions along 
the circumference of the rings. The potential felt by a particle located 
at position $x$ can be expressed by substituting the azimuthal 
angle by $\theta = 2\pi x / d$, where $d=2\pi R$ is the circumference 
of the ring, leading to 
$U(\tau,x) = -gmR \sin\tau + 2gmR\sin\tau \sin^2(\pi x/d)$.
The resulting tilting potential can be 
conceptualized~\cite{ringlattice2019} as a periodic lattice 
potential $V_0\sin^2(\pi x/d)$ of amplitude $V_0 = 2gmR\sin\tau=2m\omega^2R^2$. In this context, the natural energy scale 
in a lattice is given by $E_L=\hbar^2\pi^2/(2md^2)=\hbar^2/(8mR^2)$, 
referred to as the {\em recoil energy}~\cite{RevModPhys.80.885}.
The dimensionless ratio between lattice amplitude and the recoil energy, denoted as 
$s=(1/2)\;V_0/E_L 
= 8m^2\omega^2R^4/\hbar^2$ 
(with an extra $1/2$ factor as compared to the usual convention~\cite{Morsch2006}
), quantifies the lattice depth and matches in absolute value the 
Mathieu-function parameter $s=|q|$. 
A shallow lattice is characterized by $s\lesssim 1$ while a deep 
lattice corresponds to $s\gg 1$. As in the tilted ring only a single 
lattice site is available, i.e. $0\leq x \leq d$, the ring displays 
a minimal band structure composed of a single Bloch state, 
with zero quasimomentum, per energy band~\cite{Kittel}.
In Figure~\ref{fig:linear}(a) two opposite limits are shown, corresponding 
to  shallow ($s=0.5$) and  deep ($s=32$) lattices. 
The three lowest-energy eigenstates $\psi_{n,0}$ are shown 
with $n=0,1,2,\dots,$ standing for the band index and reflecting 
the number of nodes. As expected, the density becomes increasingly 
localized as the lattice amplitude is augmented. In the limit of 
very large $s$, the lowest eigenstates are similar to those of a 
harmonic oscillator.
	
For the interacting gas, the system dynamics obeys the 
time-dependent Gross-Pitaevskii equation
\begin{equation}
i\hbar{\partial_t\psi}=
\left(-\frac{\hbar^2}{2mR^2}\partial_\theta^2 -m\omega^2 R^2\cos\theta +\gamma|\psi|^2\right)\,  \psi,
\label{eq:GP1D}
\end{equation}
where $\gamma=-2\hbar^2/(ma_{1D})<0$ denotes the coupling constant 
of the attractive interparticle interaction and $a_{1D}$ is the 
$s$-wave scattering length \cite{Olshanii1998}. The number of particles fixes the 
wave function normalization $N=R\int d\theta\,|\psi|^2$, while the 
energy $E$ is provided by the functional ${E[\psi]} = R\int d\theta\left({\hbar^2 |\partial_\theta\psi|^2}/{2mR^2}-m\omega^2R^2\cos\theta\,|\psi|^2+{\gamma}|\psi|^4/2\right)$. 
Expectation values are computed as 
quantum-mechanical averages, 
$\langle A\rangle =(1/N)R\int d\theta\, \psi^* A \psi$, where the $1/N$ appears due to the wave function normalization.
Figure~\ref{fig:oscillations}(b) illustrates 
the effect of increasing attractive interactions 
on the ground-state density profile, which 
becomes increasingly localized. When the interparticle interaction dominates over the external 
potential, the ground state approaches a free bright soliton centered at $\theta_0=0$,
\begin{align}
\psi_{BS}(\theta)=\sqrt{\frac{N}{2\xi}}\,\mbox{sech}\left(R\frac{\theta-\theta_0}{\xi}\right),
\label{eq:BS}
\end{align}  
whose width $\xi=2\hbar^2/(m|\gamma|N) = a_{1D}/N$ scales inversely with the number 
of particles, while the peak density does quadratically. 

The close analogy between this system and the classical pendulum is evidenced by the time evolution of expectation values, as obtained from the Ehrenfest equations (see Appendix~\ref{sec:methods}). 
An illustrative example is represented in Fig.~\ref{fig:oscillations}(c), where the evolution of the center of mass, $\langle\theta\rangle$, and the angular momentum, $\langle \hat p_\theta \rangle$, is shown for a soliton of width $\xi=R/4$ in a deep lattice with $s=32$.
The time evolution follows the equations of motion $m R^2\,d \langle\theta\rangle/dt=\langle \hat p_\theta \rangle$ and $d \langle \hat p_\theta\rangle/dt=-\langle \partial_\theta U\rangle$ and, in agreement with Kohn's theorem, is independent of the interaction strength.
Hence, the nonlinear equation of the pendulum is $d^2\langle\theta\rangle/dt^2+\omega^2\langle\sin\theta\rangle=0$. 
Nevertheless, the analogy with the classical system is not always fulfilled. Had we chosen a wider wave packet (say $\xi=2 R/5$) centered at the potential maximum as the initial state, it would spread over the ring, since the wave packet falls onto both sides of the maximum, as illustrated in Fig.~\ref{fig:linear}(b). We show next that this observation reflects the lack of the classical equilibrium point at $\theta_0=\pi$, whose existence depends on the system parameters.	
	
{\bf Nonlinear spectrum.} 
The lattice viewpoint offers valuable insight into the structure of the spectrum of the nonlinear system. 
By switching on the interaction, the linear Bloch states~\cite{Kittel} characterized by a given quasimomentum, extend into the nonlinear regime while maintaining the number of nodes, which stands as a distinctive topological feature. In addition, nonlinear lattice systems allow for new stationary states of different nature, namely the {\em gap solitons}. Contrary to Bloch 
waves in an infinite lattice, which are extended states over the whole system, gap solitons are localized states that occupy a few lattice sites (see Ref.~\cite{Morsch2006}, and references herein). In static systems, such states emerge within the energy gaps of the underlying linear system beyond an interaction threshold~\cite{Louis2003, ringlattice2019}. In the tilted ring, gap solitons emerge as states centered at the potential maximum. 

\begin{figure}[tb]
\centering
\includegraphics[width=\linewidth]{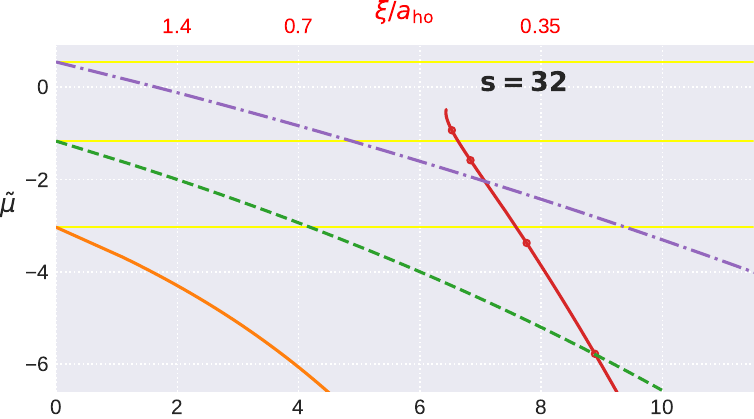}\\
\vspace{0.25cm}
\includegraphics[width=\linewidth]{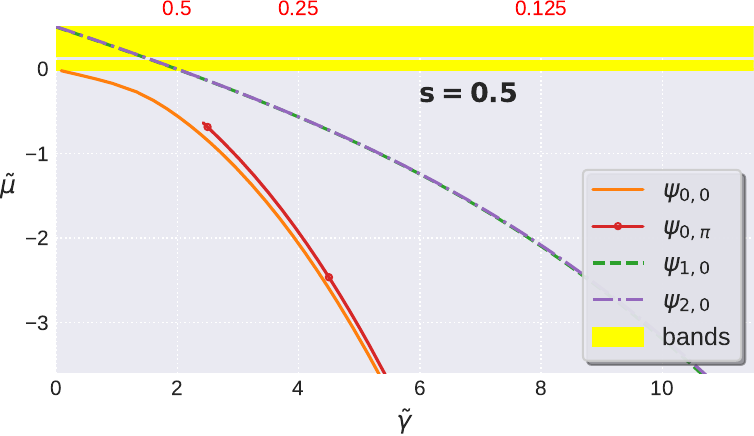}
\caption{
Chemical potential of stationary states $\tilde \mu=\mu/(\hbar^2/mR^2)$ versus interaction strength $\tilde \gamma= |\gamma| m N R/\hbar^2$, for (a) deep potential $s=32$, and (b) shallow potential $s=0.5$. The linear energy bands of an infinite lattice with the same spacing (yellow shaded regions) are shown as a reference. The top horizontal axis of each graph indicates the width of a free soliton $\xi$ in units of the harmonic oscillator length $a_{\rm ho}$. }
\label{fig:mu_N}
\end{figure}

Figure~\ref{fig:mu_N} illustrates the characteristic differences in the spectrum between shallow ($s=0.5$) and deep ($s=32$) lattices.
The chemical potential is reported as a function of the interaction strength.
The states labeled by $\psi_{n,0}$ represent the nonlinear continuation of the linear solutions shown in Fig.~\ref{fig:linear}. 
These states have the center of mass located at the potential minimum $\theta=0$. 
Beyond a certain interaction threshold, a new nonlinear state emerges (shown by the continuous line with symbols and labeled as $\psi_{0,\pi}$) with location at the potential maximum.  
Thus, these states are centered in the classically-unstable equilibrium position of the inverted pendulum with $\theta=\pi$. 

The emergence of equilibrium states at the potential maximum can be understood by expanding the potential around $\theta=\pi$, up to second order in the position.
This results in a expulsive harmonic potential $U(\pi+\delta\theta)-U(\pi)\approx -m\omega^2 R^2\delta\theta^2/2$. 
Here, the soliton Eq.~(\ref{eq:BS}) centered in $\theta_0=\pi$, with the soliton width $\xi$ as variational parameter, 
provides an ansatz. 
The energy in this system is given by the standard Gross-Pitaevskii energy functional (see, for example, Ref.~~\cite{pitaevskii_bose-einstein_2003} for details).
By performing the energy minimization we arrive to the following quartic equation $1-(a_{ho}/\xi_f)\xi/a_{ho}+(\pi/4)(\xi/a_{ho})^4=0$, where $\xi_f=2\hbar^2/(m|\gamma| N)$ is the free soliton width.
This expression is valid when both $\xi\ll R$ and $\xi_f\ll R$. 
For a wide soliton with $\xi\gg a_{ho}$, no real solution exists, indicating an absence of minimum in the energy functional and a lack of a stationary state in the vicinity of the potential maximum. 
On the contrary, for a narrow soliton with $\xi\ll a_{ho}$, the stationary solution exists with its width being roughly the same as that of a free soliton $\xi\approx\xi_f$. 
The stationary solution exists when the soliton width is smaller than the threshold value $\xi_f/a_{ho}=(0.75/\pi^{1/3})^{0.75}\approx 0.6$, or equivalently, for interaction strengths larger than $|\gamma| N\approx 3.3 \sqrt{\hbar^3\omega/m}$. 
These values provide a reasonable estimate of the threshold value, $\xi_f/a_{ho}\approx 0.45$, shown in Fig.~\ref{fig:mu_N}.
	
\begin{figure}[tb]
\flushleft ({\bf a})\\
\centering
\includegraphics[width=0.95\linewidth]{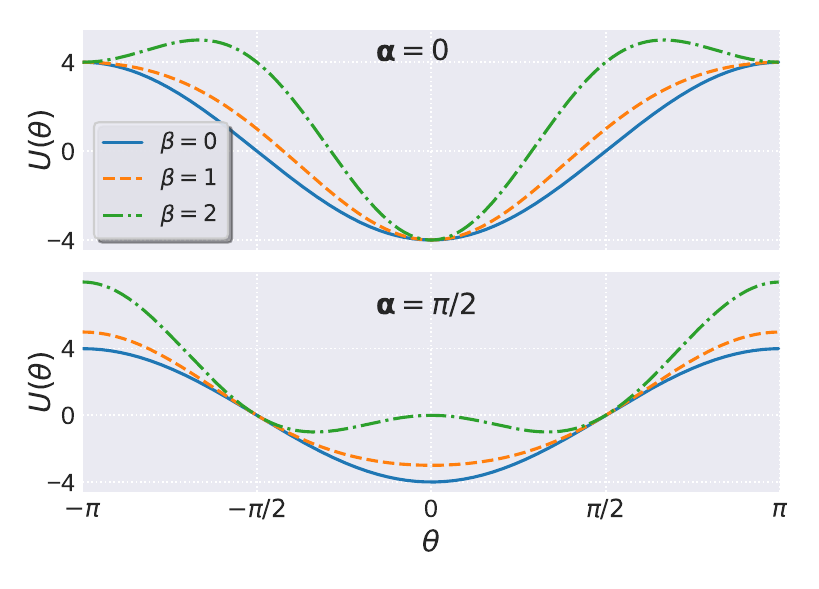}\\  \vspace{-0.5cm}
\flushleft ({\bf b})\\
\centering
\includegraphics[width=\linewidth]{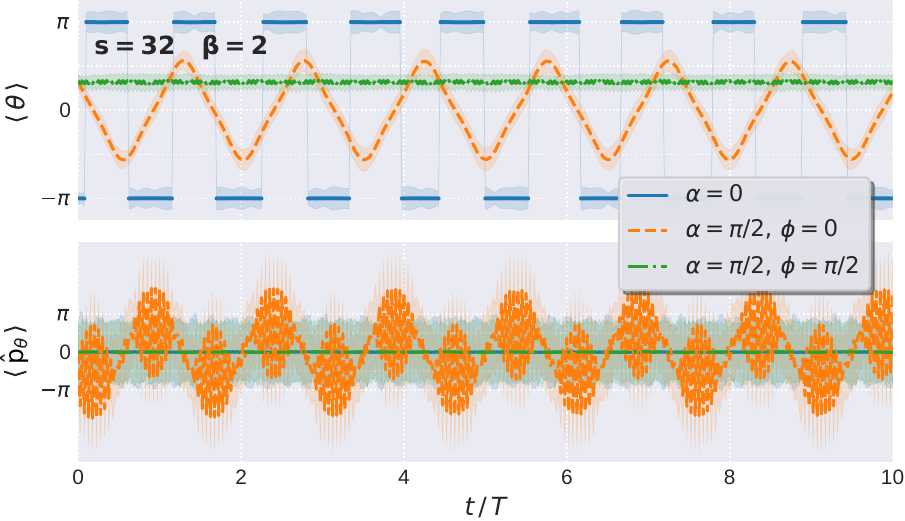}
		
\caption{
(a) Effective classical potential in the presence of a fast pivot vibration along the vertical (top panel) and horizontal (bottom panel) directions. 
(b) Time evolution of wave packets with same parameters as in Fig.~\ref{fig:oscillations}(c), centered at equilibrium points of the effective potential ($\theta_0=\pi$ at $\alpha=0$, solid line, and $\theta_0=\pi/3$ at $\alpha=\pi/2$, discontinuous lines) and subject to a pivot vibration of amplitude $\ell=0.1\,R$; the phase factor $\phi$ of the fast vibration imprints a phase profile on the initial state and controls the particle current (negative for $\phi=0$, dashed line, and zero for $\phi=\pi/2$, dot-dashed line). }
\label{fig:Ueff}
\end{figure}

{\bf Dynamical stabilization.} 
The classical pendulum possesses only a single stable position, corresponding to the pendulum stabilized at the bottom, $\theta_1=0$, while the equilibrium of the pendulum in the top position, $\theta_2=\pi$, is dynamically unstable. 
Figure~\ref{fig:oscillations}(c) illustrates that a similar phenomenon occurs in the tilted ring: a weak perturbation, generated by adding a sinusoidal wave on the initial state that shifts the center of mass to $\theta_0=-3.139$, makes the soliton roll down. 
Fascinatingly, the inverted position of the pendulum can be stabilized by inducing fast vertical vibrations (along the gravitational field direction) of the pendulum pivot as experimentally shown and mathematically proved by introducing the method of averaging of the fast variables by Pyotr Kapitza in his influential paper~\cite{Kapitza1951}. 
Commonly, the driven pendulum is referred to as a Kapitza pendulum\cite{astrakharchik2011numerical}. 
In contrast, the horizontal vibration moves the equilibrium point to a new position $\theta_2\neq\pi$ (see Ref.~\cite{Landau1960}, $\S  5,\,\S 30$). 
The potential energy of the pivot vibration of frequency $\Omega\gg\omega$ along a generic angle $\alpha$ can be calculated as the work $W= -{\bf f}\cdot{\bf r}$ done by the 2D oscillating force ${\bf f}=(\cos\alpha,\,\sin\alpha) \,m\Omega^2 \ell\,\sin(\Omega\,t)$ acting within the ring plane, where $\ell$ $\ll R$ is the characteristic amplitude of the vibration, and ${\bf r}=R(\cos\theta,\,\sin\theta)$ are the ring coordinates. 
As a result, the total potential energy becomes
\begin{align}
U(\theta,t)=-m\omega^2 R^2\left\lbrace\left[1-\frac{\Omega^2 \ell}{\omega^2 R}\sin(\Omega t)\cos\alpha\right]\cos\theta \nonumber \right. \\
\left. -\frac{\Omega^2 \ell}{\omega^2 R}\sin(\Omega t)\sin\alpha\,\sin\theta\right\rbrace .
\label{eq:potential}
\end{align} 
The idea of Kapitza is based on separating processes occurring at slow and fast time scales, i.e. corresponding to low ($\omega$) and high ($\Omega$) frequencies. 
Then, the slow motion is governed by the effective time-independent potential obtained by time averaging over the fast oscillations,
\begin{align}
U_{\rm eff}(\theta)=-m\omega^2 R^2\left\lbrace\left[\cos\theta-\left(\frac{\beta\cos\alpha\sin\theta}{2}\right)^2\right] \right.\nonumber\\ \left.
-\left(\frac{\beta\sin\alpha\cos\theta}{2}\right)^2\right\rbrace .
\label{eq:Ueff}
\end{align}
where $\beta={\Omega \ell}/({\omega R})$ is the ratio between the velocity that characterizes the vibration, $\Omega \ell$, and a velocity associated with the unperturbed pendulum, $\omega R$. The transition to (classical) stable states takes place at $\beta=\sqrt{2}$. 
The stability criterion can be stated in terms of energetic quantities, i.e. that the kinetic energy, induced by the driven oscillations, should be large compared to the potential energy above the pivotal point. 
Figure~\ref{fig:Ueff}(a) presents characteristic examples illustrating how the effective potential depends on the driving strength $\beta$ in a deep lattice (specifically, we use $s=32$).
In Fig.~\ref{fig:Ueff}(a) we focus on the case of vertical driving, i.e. $\alpha=0$.
While for weak driving vibrations ($\beta=0$ and $\beta=1$) only a single minimum exists, corresponding to the usual bottom position of the pendulum, $\theta = 0$, for strong driving ($\beta=2$) the pendulum might flip and a second stable minimum is formed for $\theta = \pm \pi$, corresponding to an inverted pendulum.
The case of horizontal driving, $\alpha=\pi/2$, is illustrated in Fig.~\ref{fig:Ueff}(b).
For weak driving oscillations, the usual single minimum is observed. Instead, for sufficiently strong driving, the new (classical)  minima appear at $\theta_{1,2}=\pm\pi/3$, while $\theta=0$ becomes a local maximum. 
	
We have tested the stability of quantum states [a wave packet similar to the one described in Fig.~\ref{fig:oscillations}(c)] subject to the time-dependent potential Eq.~(\ref{eq:potential}) with vibration amplitude $\ell=0.1\,R$, and centered at the classical equilibrium points of Eq.~(\ref{eq:Ueff}); the outcome is presented in Fig.~\ref{fig:Ueff}(b). In all considered cases, oscillations of frequency $\Omega$ can be  observed as fast beating in the evolution of average momentum. For $\alpha=0$ and $\beta=2$ (that is $\Omega=20\,\omega$) the equilibrium point $\theta_2=\pi$ is stable (solid blue curve), 
whereas for $\beta=1$ (not shown) it is not, in agreement with the classical prediction. The situation is not that clear for $\alpha=\pi/2$, since $\beta=2$ (and $\phi=0$, see below) does not lead to a static state; instead, the state is induced to tunnel through the local maximum separating the two minima (dashed line), and higher vibrations (for instance $\beta=3$, not shown) produce only a partial self trapping.
However, contrary to the classical case, the phase of the vibration is key for the quantum pendulum; introducing a phase factor $\phi$ in Eq.~(\ref{eq:potential}) by the substitution $\Omega t\rightarrow \Omega t+\phi$ has a crucial influence on stability: although $\phi=0$ does not reproduce the classical result, $\phi=\pi/2$ does (dash-dotted line). The cause resides in the velocity of the initial state (or equivalently its phase) induced by the fast oscillations and controlled by this phase factor, by virtue of which $\phi=\pi/2$ produces a zero velocity. This mechanism is not qualitatively distinct from the usual phase imprinting technique employed in ultracold-gas experiments~\cite{Dobrek1999}, and stands as an additional, key stabilization effect in quantum systems, along with the effective classical potential Eq.~(\ref{eq:Ueff}), of Kapitza's procedure.	
	
{\bf Measuring gravity.} Historically, pendulums were the first devices used to measure gravity, both for absolute and relative measurements, and also the most accurate ones until the second half of the twentieth century, reaching values of $\delta g\approx 10^{-6}g$; afterwards, they were replaced by free-fall apparatus~\cite{Zumberge2021}. Modern versions of the latter are made of ultracold atoms~\cite{Debs2011}, and have reached top-level performance $\delta g\approx 10^{-9}g$~\cite{Menoret2018,Stray2022}. By means of atomic interferometry, based on a sequence of Raman pulses that split and reunite the falling atomic clouds~\cite{Kasevich1991}, high-contrast fringes are produced that provide the acceleration of gravity. 

Differently, pendulum gravimetry relies on measuring the oscillation period, hence length, positions and corresponding times have to be tracked over repeated, small oscillations. In the present quantum pendulum, where dissipation can be ruled out, the atomic cloud can be strongly localized; for instance, an atomic cloud of $N=2.5\times 10^4$ particles of $^7$Li with interaction strength $\gamma=-2\pi \nu \hbar a_0 $, where $a_0$ is the Bohr radius, and harmonic transverse confinement of frequency $\nu=350$~Hz [similar parameters as in the experiment of Ref.~\cite{Nguyen2017}, and away from the critical particle number for collapse conditions $N_c\approx 4.5\times 10^4 $ see Refs.~\cite{PerezGarcia1997,PerezGarcia1998}], has a typical size of $\xi\approx 0.6\,\mu$m (doubling $\nu$ reduces $\xi$ by half and $N_c$ by $\sqrt{2}$). Even within a small ring of radius $R=12\,\mu$m, as those of Ref.~\cite{Beattie2013}, the relevant parameter $R/\xi=20$ is large enough to closely reproduce the dynamics of a classical pendulum (with expected relative differences of $\delta\sim5\times 10^{-4}$ in the motional periods, see Appendix~\ref{sec:methods}). In this case, assuming that shining far-from-resonance light perpendicular to the ring on the atomic cloud does not modify the dynamics, as for classical particles, laser photo-detection could be used to track the motion with expected accuracy of, at least, the order or a few percent (as typically reported in oscillation measurements~\cite{Fang2014,Huang2019}); in-trap non-destructive absorption imaging~\cite{Nguyen2017} could also be performed. This uncertainty in measuring the motional period might be reduced by resorting to similar procedures as for the classical reversible pendulums of Kater and Bessel~\cite{Poynting1908}: the measurement of two close but different periods. In the ring, they are associated with different tilts, or equivalently to different lengths $L_j=R/\sin\tau_j$, for $j=1,\,2$, so that gravity is obtained from $g=(2\pi)^2\,(L_1^2-L_2^2)/(L_1T_1^2-L_2T_2^2)$. 
	
A precise determination of the period would also allow one to use the tilted ring for performing sensitive measurements of gravity~\cite{Stray2022}. By initially preparing the BEC at the energy maximum $\theta=\pi$, where small displacements produce large differences in the period, the presence of a sensible mass at the surface, horizontally located with respect to the ring axis such that $\delta g$ and $g$ could act orthogonally, would induce an angular displacement $\delta\theta\approx\tan\delta\theta=\delta g/g$. This translates into pendulum periods [see Appendix~\ref{sec:methods}, Eq.~(\ref{eq:T})] that 
can be approximated by $T_k(\delta\theta)\approx (2/\pi)(3 \ln2-\ln \delta\theta)\,T$; for instance, for a void of mass of 100~kg at 1-m apart one obtains $\delta g/g\approx 7\times 10^{-10}$ and $T_k=145.5\,T$, while for twice that mass, $\delta g/g\approx 14\times 10^{-10}$, the period is $T_k=141.2\,T$.

	\appendix
	\section{}
	\label{sec:methods}

\subsection{ Numerical solutions}
 Equation~(\ref{eq:GP1D}) has been numerically solved through FFT techniques for the spatial discretization, and standard, high-order (typically 5 to 9) time integrators of Julia programming language. For time-independent solutions, both imaginary time evolution, for the ground state, and Newton method, for excited states as those presented in Fig. \ref{fig:mu_N}, have been used.
	
\subsection{ Adiabatic switch of fast vibrations}
 Instead of suddenly turning on the fast vibrations, as described in the main text, the dynamical stabilization of equilibrium points can also be performed in an adiabatic way. To this end, we have chosen a setup with similar parameters as the ones used in Fig.~\ref{fig:Ueff}(b) and $\alpha=\pi/2$, but with an initial soliton state of varying amplitude (according to $\tilde\gamma=8$ or 16), centered at the equilibrium point $\theta=0$ of the non-vibrated system. Subsequently, the amplitude and high frequency of the vibrating potential are adiabatically ramped up, $U(\Omega)=f(t)\,m \Omega^2 \,\ell R\,\sin[f(t)\,\Omega t+\phi]\,\sin\theta$, where $f(t)=\tanh(0.1\,\omega\,t)$. {Figure~\ref{fig:Adiabatic} shows some characteristic examples for the time evolution of the center of mass in the adiabatic case. 
As it can be inferred from this figure,
the phase factor $\phi$ still plays a relevant role, and eventually, for $\phi=\pi/2$, the initial state rolls downhill in the effective potential generated by the fast vibration towards one of its energy minima at $\theta=\pi/3$; higher and less regular oscillations around the energy minimum are observed for the more localized soliton (at $\tilde\gamma=16$) due to the higher momentum uncertainty.
\begin{figure}[tb]
\centering
\includegraphics[width=1\linewidth]{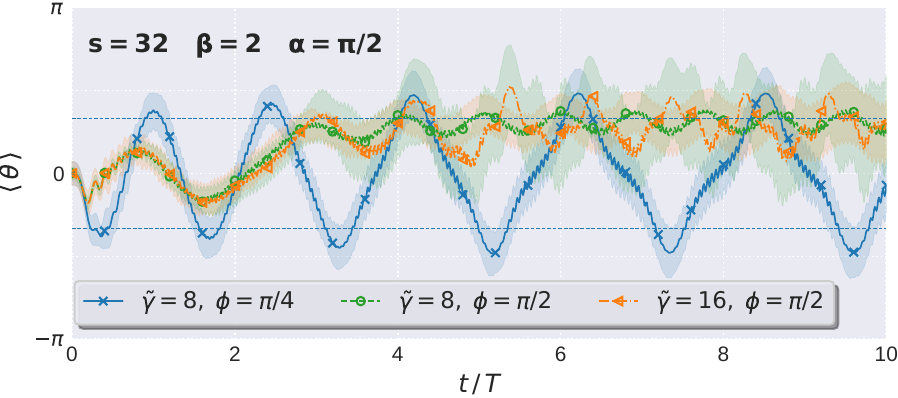}
\caption{Time evolution of the center of mass position in the case of an adiabatic switch of the fast vibration along the direction $\alpha=\pi/2$.
The tilted ring is characterized by a potential depth of $s=32$. 
The initial soliton states, featured by the interaction $\tilde \gamma$ and situated at $\theta=0$, move towards the effective energy minima (indicated by horizontal, thin dashed lines) depending on the vibration phase $\phi$ (see text for additional details).}
\label{fig:Adiabatic}
\end{figure}	

\subsection{ Classical vs quantum periods of oscillation}
For oscillatory motion, the classical equation of the pendulum $\ddot \theta+\omega^2\sin\theta=0$ has general solutions in terms of elliptic functions $\theta(t)\sim \arcsin\left[k\,\mbox{sn}(\omega t,k)\right]$ (see, for instance Refs.~\cite{Symon,Tabor}), where $\mbox{sn}(\omega t,k)$ is the Jacobi elliptic sine function~\cite{NIST:DLMF} of modulus $k=\sin(\theta_m/2)$, with $\theta_m$ being the maximum, turning-point angle. For generic initial conditions, $\theta(0)=\theta_0$ and $\dot \theta(0)=\dot\theta_0$, the solution reads
\begin{align}
\theta(t)=\theta_0+&2 \left\lbrace\, \arcsin\left[k\,\mbox{sn}(\omega t+\varphi_0,k)\right] \right. \nonumber\\ & \left.\qquad - \arcsin\left[k\,\mbox{sn}(\varphi_0,k)\right] \,\right\rbrace,
\label{eq:angle}
\end{align}
from which, it follows the angular velocity 
\begin{align}
\dot\theta(t)=2 k \omega\, \mbox{cn}(\omega t+\varphi_0,k),
\label{eq:dot_angle}
\end{align} 
where $\mbox{cn}$ is the Jacobi elliptic cosine function, hence $\varphi_0=\mbox{arc\,cn}[\dot\theta_0/(2k\,\omega),\,k]$. The pendulum period is given in terms of the complete elliptic integral of the first kind $K(k)$~\cite{NIST:DLMF} as
\begin{align}
T_k=\frac{K(k)}{\pi/2}\, T,
\label{eq:T}
\end{align}
such that it differs from the period $T=2\pi/\omega$, achieved in the approximation of small oscillations, by the factor $K(k)/(\pi/2)\in [1,\,\infty]$ (a monotonically increasing function of $k$) for $k\in  [0,\,1]$, that is for $\theta_m\in[0,\,\pi]$.
\begin{figure}[tb]
\flushleft ({\bf a})\\
\centering
\includegraphics[width=1\linewidth]{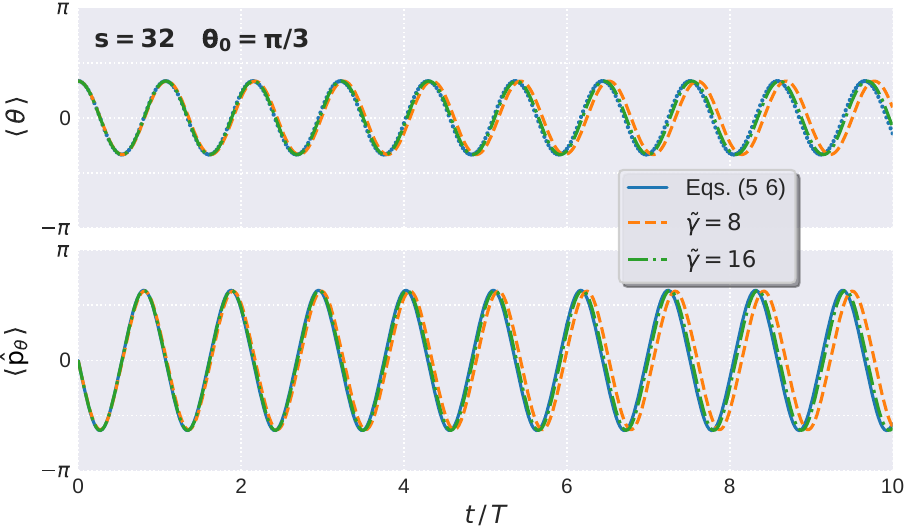}\\  \vspace{-0.2cm}
\flushleft ({\bf b})\hspace{7.6cm} ({\bf c})\\
\includegraphics[width=0.49\linewidth]{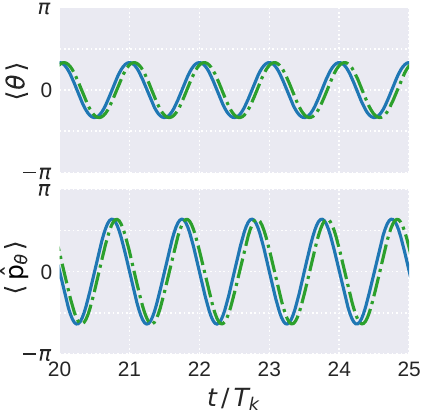} 
\includegraphics[width=0.49\linewidth]{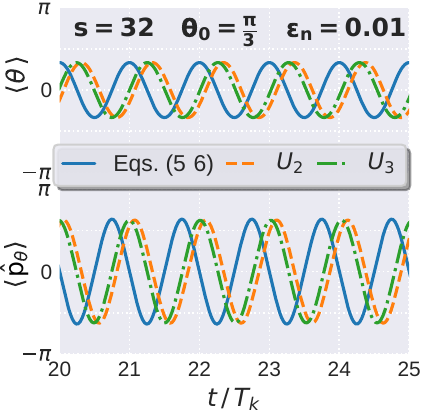} \hspace{8mm}
\caption{
Comparison between the motions of the classical pendulum and the soliton as given by Eq.~(\ref{eq:BS}). 
Time evolution of the center of mass, $\langle \theta\rangle$ and the angular momentum, $\langle \hat p_\theta \rangle$ is shown in a tilted ring characterized by $s=32$ using initial conditions $\theta_0=\pi/3$ and $\dot\theta_0=0$. Different interactions, parameterized by $\tilde \gamma=2R/\xi$, translate into different wave-packet widths $\xi$.
(a) Short-time evolution for $\tilde \gamma=8$ and $\tilde \gamma=16$ (b) Long-time evolution for $\tilde \gamma=16$. (c) Time evolution in a perturbed ring potential $U_n(\theta)=-m\omega^2 R^2[\cos\theta-\epsilon_n\cos(n\,\theta)]$, with $n=2,\,3$ . }
\label{fig:Tcomp}
\end{figure}

Figure~\ref{fig:Tcomp} shows a comparison between the classical predictions of Eqs.~(\ref{eq:angle}-\ref{eq:T}) and the mean values of the motion of free-soliton wave packets~\ref{eq:BS} in a tilted ring. As one could expect, the period predicted by Eq.~(\ref{eq:T}), $T_k=1.0732\,T$, is better approached at higher interactions $\tilde\gamma=16$. By averaging over 25 periods of oscillation in order to minimize the uncertainty in the measured times, we obtain the differences $\delta(\tilde\gamma)=[T_k(\tilde{\gamma})-T_k]/T_k$ to be $\delta(\tilde{\gamma}=16)=3.36\times 10^{-3}$, and $\delta(\tilde{\gamma}=8)=1.23\times 10^{-2}$; a slight reduction is found for initial conditions closer to equilibrium positions, for instance, $\delta(\tilde{\gamma}=16)=3.14\times 10^{-3}$ for $\theta_0=3\pi/4$.

These differences can be understood by considering the Ehrenfest equation $d^2 \langle\theta\rangle/dt^2+\omega^2\langle\sin\theta\rangle=0$, which exactly matches the functional form of the classical equation, as $ d^2 \langle\theta\rangle/dt^2+\omega^2\sin\langle\theta\rangle=0$, only when the density profile of the soliton wave function becomes a Dirac delta function $|\psi_{BS}(\theta-\theta_0)|^2\rightarrow \delta(\theta-\theta_0)$; otherwise, the difference between both equations can be written as the power series
\begin{align}
\Delta=\,&\langle\sin\theta\rangle - \sin\langle\theta\rangle= \nonumber\\ & \sum^{n\ge 3}_{ n\,\mbox{\tiny odd}}\frac{ (-1)^\frac{n-1}{2}}{n!}\,
\left(\langle\theta^n\rangle-\left\langle\theta\right\rangle^n \right).
\label{eq:sintheta}
\end{align}

For a free soliton state Eq.~(\ref{eq:BS}) of width $\xi$ and moving center $\theta_0(t)$, the above series can be approximated, for large ratios $(R/\xi=\tilde \gamma/2)\gg 1$, up to second order, by 
\begin{align}
\Delta(\tilde \gamma)\approx -\frac{\pi^2}{6\,\tilde \gamma^2}\,\sin\langle\theta\rangle,
\label{eq:difference}
\end{align}
where we made use of ${\langle\theta\rangle}=\theta_0(t)$. This result amounts to having an effective classical equation with a slightly reduced angular frequency $\omega(\tilde \gamma)=\omega\,\sqrt{1-{\pi^2}/({6\,\tilde \gamma^2})}$, or, equivalently, a slightly increased small-oscillation period 
\begin{align}
T(\tilde \gamma)\approx\left(1+\frac{\pi^2}{12\,\tilde \gamma^2}\right)\, {T}\,.
\label{eq:Tlambda}
\end{align}
Evaluated at the values $\tilde \gamma=8, \, 16$, used in 
Fig.~\ref{fig:Tcomp}, this estimate produces 
$\delta(\tilde\gamma=8)=1.31\times 10^{-2}$, and
$\delta(\tilde\gamma=16)=3.23\times 10^{-3}$, in good agreement with the measured results previously reported.

{\itshape Ring roundness}.- Small azimuthal variations in the ring potential can also give rise to alterations in the period of the motion. In order to estimate the effect of such variation, we have introduced a perturbed potential $U_n(\theta)=-m\omega^2 R^2[\cos\theta-\epsilon_n\cos(n\,\theta)]$, with a small integer  $n>1$, and $\epsilon_n\ll1$. It translates into an extra force term $F_n(\theta)=\epsilon_n m\omega^2 R\,n\langle \sin(n\theta)\rangle$ in the corresponding Ehrenfest equation that can also be calculated as a power series up to second order in $\tilde\gamma^{-1}$ as
\begin{align}
F_n(\tilde \gamma)\approx  \epsilon_n\,m\omega^2 R\,n\, \left(1-\frac{n^2\pi^2}{6\,\tilde \gamma^2}\right)\,\sin\langle n\,\theta\rangle\,.
\label{eq:Fn}
\end{align}
In the limit of small oscillations, a perturbed single frequency, $\omega_n(\tilde\gamma)=\omega\sqrt{1-\epsilon_n [n^2-{n^4\pi^2}/({6\,\tilde \gamma^2})]}$, can be obtained; for instance, by setting $n=2,\,3$, and  $\epsilon_n=0.01$, this estimate provides us with the order of magnitude of the perturbed period observed in our numerical results, as shown in Fig. \ref{fig:Tcomp}(c), where we have measured $\delta_2(\tilde{\gamma}=16)=1.57\times 10^{-2}$ and $\delta_3(\tilde{\gamma}=16)=1.16\times 10^{-2}$. From the reported data, it becomes clear that the azimuthal variations of the ring potential can introduce a significant source of uncertainty in the measured observables, at least in the search for high precision measurements (see e.g. Ref.~\cite{deGoer2021}); in this regard, time-averaging optical ring potentials~\cite{Bell2016} could be useful to achieve improved trap smoothness.

	\section*{Acknowledgements}
The authors are grateful to Jean Dalibard for careful reading of a draft of this paper and offering insightful comments. We are indebted to Joan Martorell for multiple discussions, careful reading of drafts, and comprehensive help. This work has been funded by Grants No. PID2020-114626GB-I00 and PID2020-113565GB-C21 by MCIN/AEI/10.13039/5011 00011033 and "Unit of Excellence Mar\' ia de Maeztu 2020-2023" award to the Institute of Cosmos Sciences, Grant CEX2019-000918-M funded by MCIN/AEI/10.13039/501100011033. We acknowledge financial support from the Generalitat de Catalunya (Grants 2021SGR01411 and 2021SGR01095).

	\bibliography{pendulum_commphys}

\end{document}